# Nonlocal Piezoresponse of LaAlO$_3$/SrTiO$_3$ Heterostructures

Mengchen Huang[1], Feng Bi[1], Chung-Wung Bark[2], Sangwoo Ryu[2], Kwang-Hwan Cho[2], Chang-Beom Eom[2] and Jeremy Levy[1#]

[1]*Department of Physics and Astronomy, University of Pittsburgh, Pittsburgh, Pennsylvania 15260, USA;* [2]*Department of Materials Science and Engineering, University of Wisconsin-Madison, Madison, Wisconsin 53706, USA*

[#]jlevy@pitt.edu

## Abstract:
The hysteretic piezoelectric response in LaAlO$_3$/SrTiO$_3$ heterostructures can provide important insights into the mechanism for interfacial conductance and its metastability under various conditions. We have performed a variety of nonlocal piezoelectric force microscopy experiments on 3 unit cell LaAlO$_3$/SrTiO$_3$ heterostructures. A hysteretic piezoresponse is observed under various environmental and driving conditions. The hysteresis is suppressed when either the sample is placed in vacuum or the interface is electrically grounded. We present a simple physical model which can account for the observed phenomena.

The interface between perovskite oxide semiconductors $LaAlO_3$ (LAO) and $SrTiO_3$ (STO) exhibits remarkable conducting [1, 2], superconducting [3, 4], magnetic [5], and spintronic properties [6, 7] that have potential relevance for an emerging field of oxide nanoelectronics [8]. Ohtomo and Hwang first reported conductivity at this interface [1] and ascribed it to a "polar catastrophe" from the polar discontinuity between LAO and $TiO_2$-terminated STO. It was later discovered that there exists a sharply defined transition from insulating to conducting interface at a critical thickness of 4 unit cells [2]. The conductivity of the interface can be modulated using a back gate applied to the STO substrate. For 3 unit cell (u.c.) LAO/STO films, this type of gating can produce a hysteretic metal-insulator transition (MIT). At this critical thickness, the MIT can also be locally controlled using a conductive atomic force microscope (c-AFM) tip [9]. The c-AFM tip deposits positive charge which is compensated by local electron accumulation at the LAO/STO interface [8, 10]. Devices such as field-effect transistors [8], photodetectors [11], rectifying junctions [12] and single-electron transistors [13] have been created with spatial resolution of 1-2 nm.

It was first reported by Bark et al [14] that the LAO/STO system (at above the critical thickness of $LaAlO_3$) exhibits a switchable piezoelectric response that is associated with the interfacial MIT. Using piezoresponse force microscopy (PFM), they showed strong hysteretic behavior which was attributed to a dynamic field-induced ionic migration process [15].

Here we describe a series of PFM experiments on 3 u.c. LAO/STO structures designed to elucidate the mechanism of hysteretic piezoresponse behavior and its relation the underlying conductivity of the LAO/STO interface. By combining c-AFM writing with PFM detection, we can try to better understand the conditions under which PFM hysteresis is observed and understand its physical origin.

Those samples are fabricated by depositing 3 u.c. LaAlO$_3$ on a TiO$_2$-terminated (001) SrTiO$_3$ substrates using pulsed laser deposition with *in situ* high pressure reflection high energy electron diffraction (RHEED) method. Growth was at a temperature of 550°C and oxygen pressure 10$^{-3}$ mbar [16]. Electrically conducting contacts to the interface were prepared by first milling 25 nm deep trenches through the LAO layer via an Ar-ion mill and then filling them with 4 nm Ti followed by 25 nm Au bilayer electrodes via sputtering. The electrical contacts are arranged to allow a central "canvas" to be patterned using c-AFM lithography. The back surface of the STO substrate is electrically contacted using silver epoxy so that a back-gate voltage $V_{bg}$ can be applied.

Figure 1 shows the experimental setup. A conductive Pt-coated silicon probe contacts the LAO surface, and is either grounded or electrically isolated. An ac voltage $V_i = V_{ac} \cos(2\pi ft)$ is applied to two electrodes that make contact to the interface. A dc voltage $V_{bg}$ is applied to the back of the sample. The ac voltage can induce a piezoresponse that is measured through the deflection of the AFM cantilever and detected using a lock-in amplifier. The frequency $f$~300 kHz is adjusted to take advantage of resonant enhancement, and is adjusted to optimize the response before measurements are performed under a given set of experimental conditions. Because the ac voltage is applied to the conductive electrodes and not the AFM tip, the PFM response is referred to as "nonlocal". Experiments are performed in one of two configurations. For some experiments the conductive AFM tip is connected to ground, while in other configurations the probe is electrically isolated. All of the experiments described below are performed at room temperature, under varying conditions of humidity and atmospheric pressure.

### *Results*

Figure 2 shows a series of non-local PFM (NL-PFM) experiments performed on a 3 unit cell (uc) LAO/STO "canvas". Measurements are taken for four combinations of conditions: with the AFM tip either floating or grounded; and for two atmospheric conditions: (40% RH) or under vacuum ($10^{-5}$ Torr pressure). Fig. 2(d,g) show the NL-PFM amplitude and phase as a function of $V_{bg}$ for an experiment in which the tip is positioned over a region of the sample with an insulating interface (Fig. 2(a)). Significant hysteresis is observed at atmospheric conditions for both grounded and isolated AFM tips. The sign of the piezoresponse changes in the region $0\text{V} < V_{bg} < 5\text{V}$, as can be seen from the change in phase of the response. By contrast, no hysteresis is observed for the experiments performed under vacuum conditions.

A similar series of experiments is performed after "writing" [9] a 1 μm x 1 μm square region and subsequently placing the AFM tip at the center of the square (Fig. 2(b)). This writing process is achieved by electrically connecting the AFM tip to a +10 V source and scanning over the square area in a raster fashion. However, this conductive square is still electrically isolated from the conducting electrodes. The results (Fig. 2(e,h)) from the NL-PFM measurement show a strong enhancement of the piezoresponse by almost a factor of three; however, the hysteretic response is still suppressed under vacuum conditions.

In a third experiment (Fig. 2(c)), this square area is electrically connected to one of the conducting electrodes using the same writing procedure. During the connection process, the AFM tip does not come into contact with the central area; instead the edge of the conducting square is brought into contact with an "L"-shaped conductive region that is connected to the electrode. After writing, the AFM tip is again placed at the same location as for the two previous sets of experiments, and the four types of measurements are repeated. Under these conditions, the hysteretic response is absent for experiments performed both under vacuum and at atmospheric conditions (Fig. 2(f,i)).

To further explore the nature of the hysteretic PFM response, a related set of experiments was performed under vacuum conditions and the presence of ions generated from a vacuum ion gauge (Fig. 3(a)). The chamber pressure was estimated to be $<10^{-4}$ Torr based on a pressure reading prior to the beginning of the experiment. Fig. 3(b) shows the NL-PFM response versus $V_{bg}$ for vacuum conditions similar to Fig. 2(a) (i.e., no square or conductive path written). Prior to this measurement, the vacuum ion gauge is left off for 30 minutes. The ion gauge is turned on and the NL-PFM is measured after 5 minutes (Fig. 3(c)), 15 minutes (Fig. 3(d)) and 30 minutes (Fig. 3(e)). Over time, a wide hysteresis loop appears with increasing width. Then, the ion gauge is turned off. The width of the hysteresis loop decays significantly after 30 minutes (Fig. 3(f)) and is restored to the original state after 90 minutes (Fig. 3(g)).

*Discussion*

This experiment setup is different from typical "local" PFM experiments. In a local geometry, an ac+dc signal is applied to the tip and the resulting piezoelectric response is detected at the same location. The applied voltage on the tip can affect the sample in many ways. Specifically, surface charge deposition or oxygen vacancies injection is both possible, and both can possibly contribute to the electromechanical response. However, in our nonlocal geometry, the biased tip effects are avoided since no voltage is applied to the tip. Surprisingly, we still observe a large and often hysteretic PFM response. The insensitivity of the response to the electrical connectedness of the tip suggests that direct charge transfer from the tip is not responsible for the NL-PFM signal.

The experimental results are consistent with a mechanism in which the role of charged adsorbates (i.e., ions) helps to regulate the hysteretic NL-PFM response. The experimental results from Fig. 2 show that hysteresis is not observed when the experiments are performed in vacuum. Furthermore, significant hysteresis is observed under vacuum conditions when ions are generated from a source. Here we propose one possible mechanism that could produce the PFM response that is related to the LAO surface adsorbates. First-principles calculations show that $H_2O$ binds strongly with $AlO_2$ outer surface at and below room temperature, and dissociates into $H^+$ and $OH^-$ [17]. The "water-cycle" mechanism that is consistent with experiments by Bi et al [18] show that ions can be readily sourced from water under ambient conditions (with RH> 30%). Experiments by Xie et al [19] have shown that many other LAO surface adsorbates can couple to the interface electronic states.

In the absence of charged surface adsorbates, one expects a small but non-negligible contribution to the NL-PFM signal that originates from coupling of the fringe electric fields across the polar LAO layer (Fig. 4(a)). While LAO is not piezoelectric in the bulk, a thin oriented layer is expected to be piezoelectric; no hysteretic response is expected due to the LAO layer in isolation. However, the electric field can attract free ions to the LAO surface (Fig. 4(b)). Such adsorbates may have a small binding energy to the surface, and this binding energy can be greatly increased if the charge is compensated by electron accumulation at the interface. These electrons can come from remote donor sites, or from the nearby conducting electrode. The surface charge and electron can form an electrostatic bound state (indicated by $V_{surf}$ and $V_{int}$), which may lead to a hysteretic relationship between the surface charge as function of $V_{bg}$.

The presence of electrons in the STO region near the interface is expected to produce noticeable elongation of the oxygen octahedra (Fig. 4(c)), due to Jahn-Teller distortion effects.

This elongation was first reported for the LAO/STO system by Maurice et al [20]. The tetragonal distortion leads to an energy splitting of the $d_{xy}$ and $d_{xz}/d_{yz}$ Ti d states. At the interface, the $d_{xy}$ state is lower in energy; a local tetragonal distortion can therefore produce lateral confinement and charge localization. The tetrahedral distortion itself can produce a large PFM response. When the interface becomes fully conducting due to c-AFM lithography (Fig. 4(c)), the interface effectively screens the top LAO surface, and hence adsorption/desorption of ions becomes unlikely as $V_{bg}$ is varied.

In summary, we have performed a series of experiments to help elucidate the nature of hysteretic piezoelectric response of LAO/STO structures. The observed NL-PFM response and its sensitivity to ambient atmospheric conditions as well as the conductivity of the interface suggest that hysteresis comes from bound states between electrons at the interface and surface adsorbates. The piezoresponse itself most likely originates from Jahn-Teller distortions of the oxygen octahedral. The NL-PFM response, in addition to helping elucidate the mechanism for interfacial conductance at the LAO/STO interface, also provides a new mechanism for probing electronic properties and correlating them with local structural distortions.

This work is supported by NSF DMR-1104191 (Levy) and NSF DMR-0906443 (Eom). The authors thank Alexei Gruverman for helpful discussions.

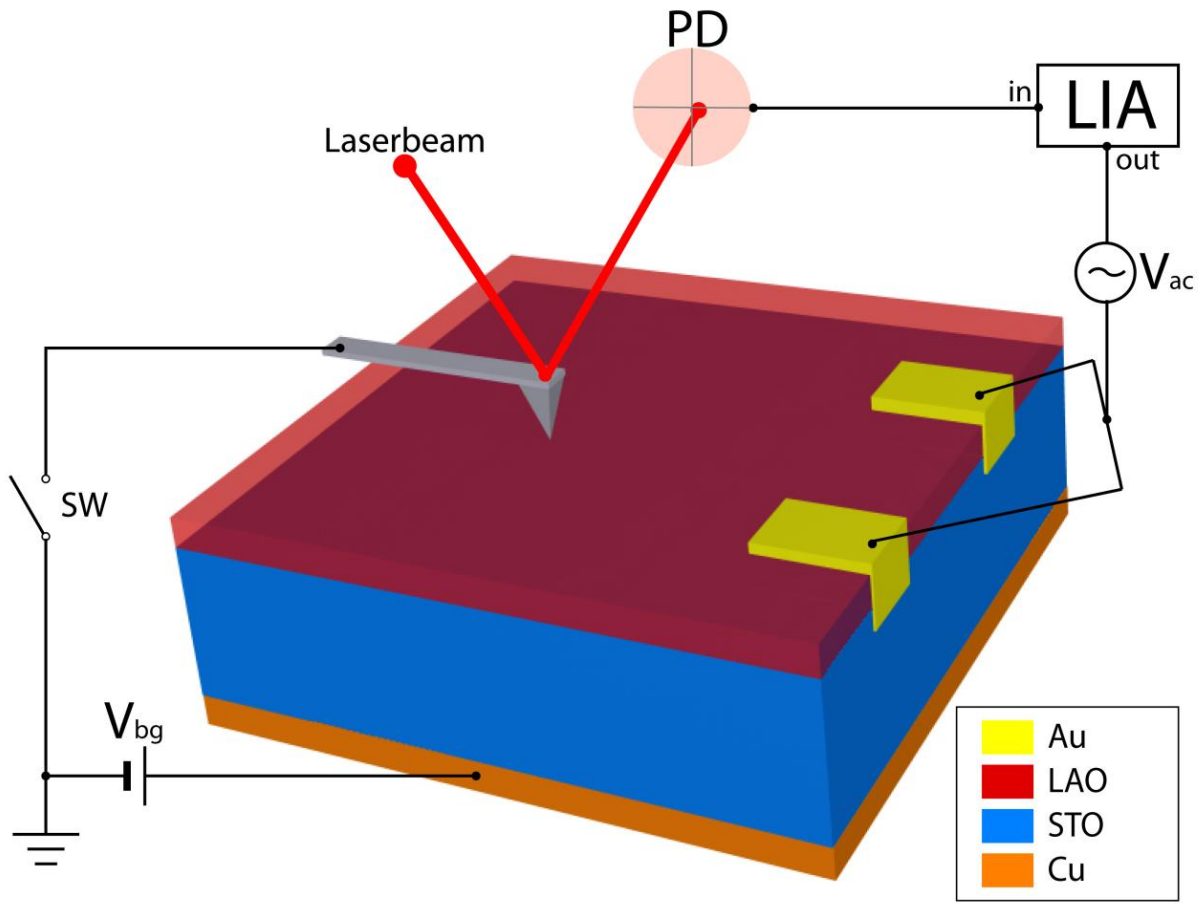

**Figure 1. Schematic of the experiment set-up.** A 3 uc LAO/STO structure is gated with Vbg at the bottom STO substrate, and at the interface by an ac voltage $V_{ac}\cos(2\pi ft)$. An AFM tip contacts the top LAO surface and the deflection is measured with a lock-in amplifier (LIA). A switch (SW) allows experiments to be performed with the AFM tip electrically grounded or isolated. The experiment can also be performed under atmospheric conditions (760 Torr, 40% RH) or under vacuum (~$10^{-5}$ Torr).

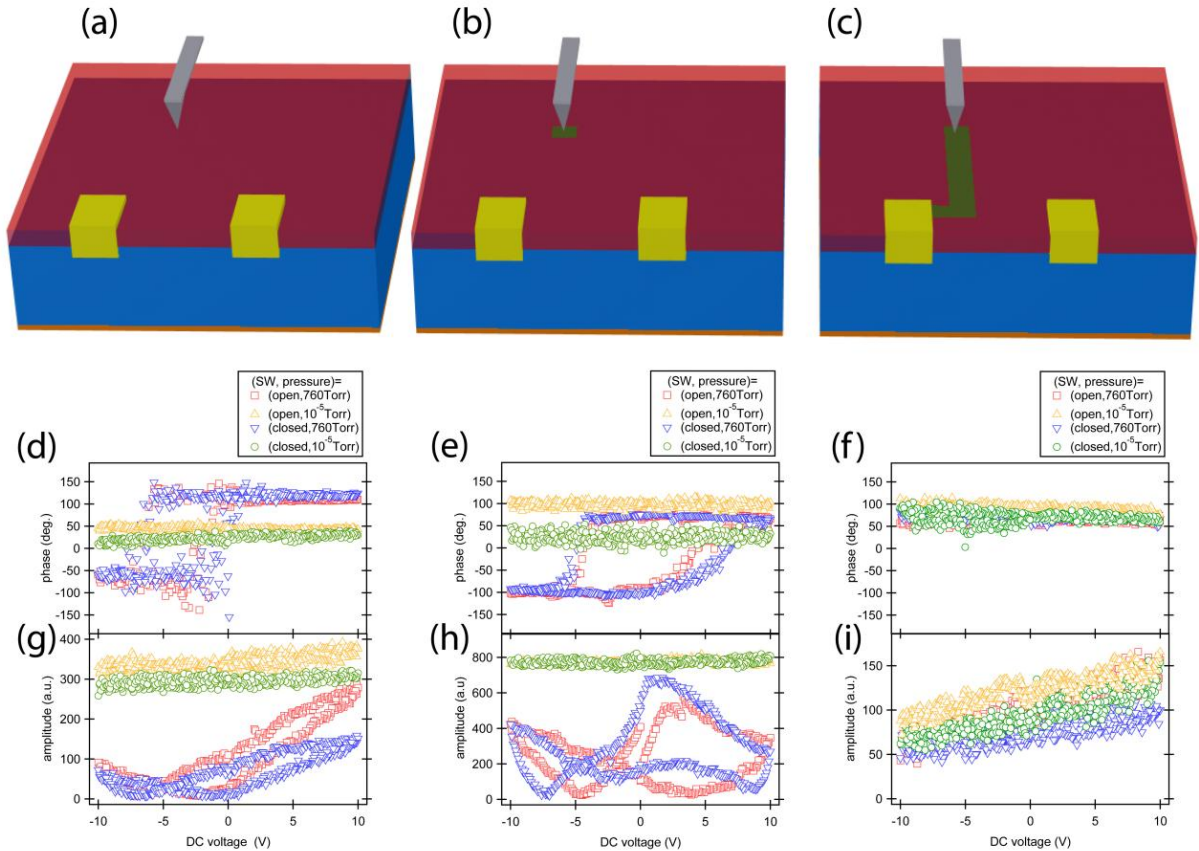

**Figure 2. Nonlocal PFM experiments.** (a) AFM tip is engaged a distance of 10 μm from the interface electrodes. (b) AFM tip is engaged after "writing" a 1um square region. (c) AFM tip is engaged after writing an "L-shaped" strip that makes contact to one of the interface electrodes. Experiments are performed under four different conditions (switch SW open or closed; pressure either 760 Torr or $10^{-5}$ Torr). (d, g) magnitude and phase for configuration (a). (e,h) magnitude and phase for configuration (b). (f,i) magnitude and phase for configuration (c).

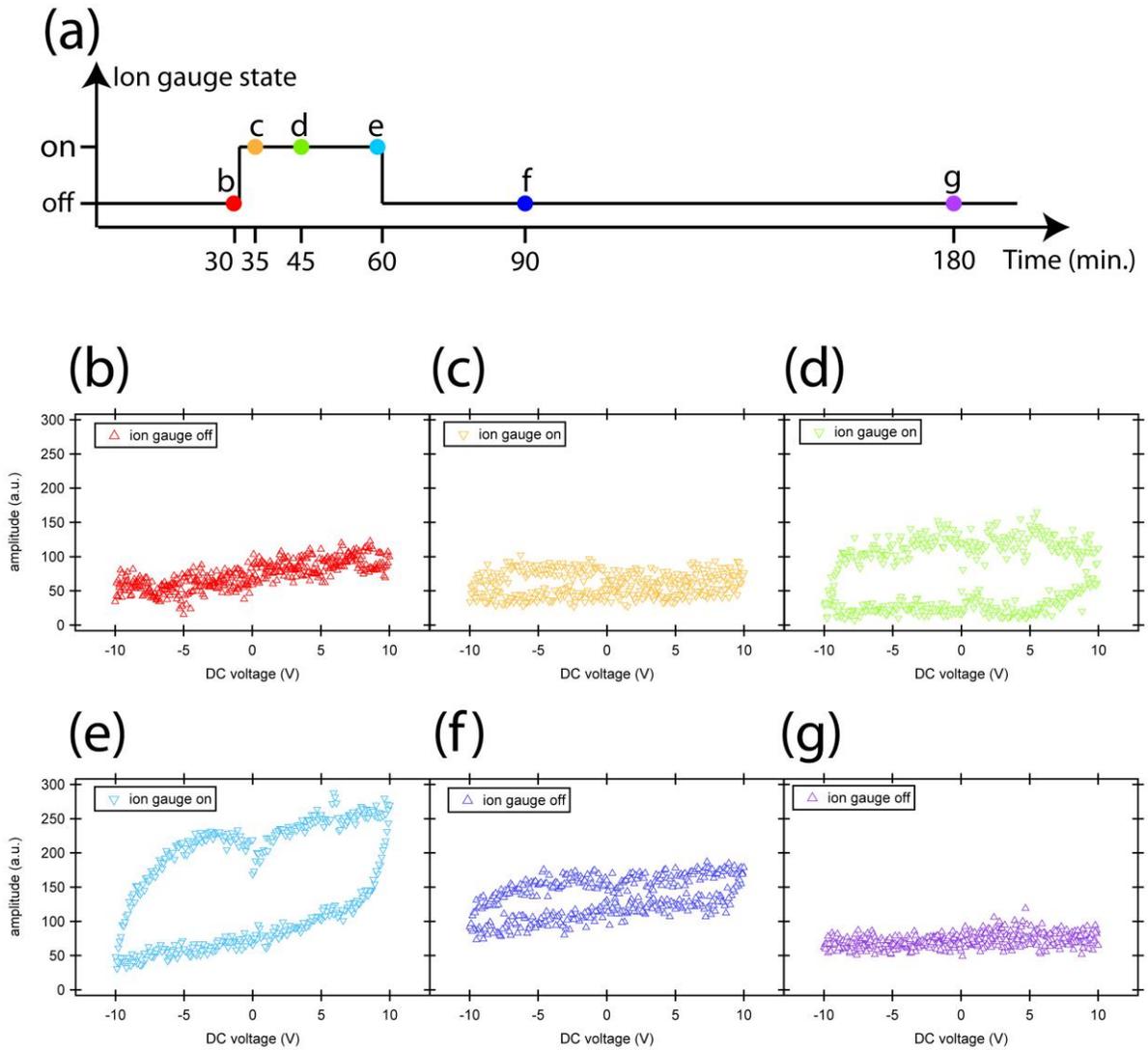

**Figure 3. NL-PFM experiments performed with ion source.** NL-PFM experiments are performed as a function of time with the ion gauge left in either the "on" or "off" state. (a) Diagram indicating the state of the ion gauge. (b-g) NL-PFM measurements performed as a function of time.

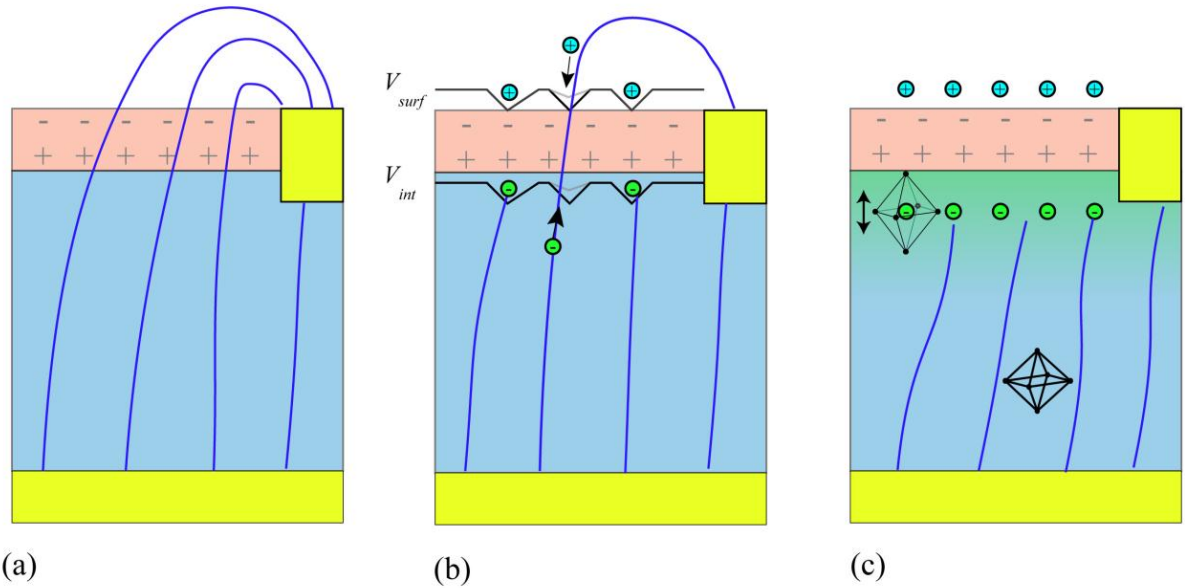

**Figure 4. Schematic of LAO/STO interface.** (a) Electric field lines (blue) generated between bottom electrode and interface contact. (b) Surface charges and interface electrons form bound states. (c) A conducting LAO/STO interface results in a Jahn-Teller distortion of the oxygen octahedral, which enhances the PFM response. The electric field from the back gate is screened, and hysteresis is not observed.